\documentclass[multphys,vecphys]{svmult}

\usepackage{makeidx}     % allows index generation
\usepackage{graphicx}    % standard LaTeX graphics tool
\usepackage{multicol}    % used for the two-column index
\usepackage{cite}

\makeindex             % used for the subject index
\begin{document}

\title{Quantum Phase Transitions of Quasi-One-Dimensional
Heisenberg Antiferromagnets}
\titlerunning{Quasi-One-Dimensional Heisenberg Antiferromagnets}

\author{Munehisa Matsumoto\inst{1}
\and Synge Todo\inst{2}
\and Chitoshi Yasuda\inst{3}
\and Hajime Takayama\inst{1}}
\authorrunning{Munehisa Matsumoto et al.}

\institute{Institute for Solid State Physics,
University of Tokyo, Chiba 277-8581, Japan
\and Department of Applied Physics,
University of Tokyo, Tokyo 113-8656, Japan
\and Computational Materials Science Center,
National Institute for Materials Science, Tsukuba 305-0047, Japan}

\maketitle

We study the ground-state phase transitions
of quasi-one-dimensional
quantum Heisenberg antiferromagnets
by the quantum Monte 
Carlo method with the
continuous-time loop algorithm
and finite-size scaling.
For a model which consists of $S=1$ chains with
bond alternation coupled
on a square lattice, we determine
the ground state phase diagram and
the universality class of the quantum phase transitions.

\section{Why Quasi-One-Dimensional Systems?}

Low-dimensional quantum antiferromagnets
have attracted much attention in recent years.
Due to quantum fluctuations,
they often have non-trivial ground states.
We investigate the ground state
of quasi-one-dimensional (Q1D)
Heisenberg antiferromagnets (HAF's)
which consist of coupled one-dimensional (1D)
spin chains with bond alternation on a square lattice.
An isolated 1D uniform spin chain has no long-range
order and a striking phenomenon associated with
the Haldane gap~\cite{haldane} is known,
namely, chains with integer spins have
a finite excitation gap over their ground states,
whereas those with half-odd-integer spins do not.
On the other hand,
HAF's on a spatially isotropic square lattice
have the long-range N\'{e}el order
in the ground state\cite{kubo}.
What kind of ground states do the models have that lie
in the intermediate region between genuine 1D systems
and two-dimensional (2D) systems?
This is our question.

\begin{figure}
\centering
\parbox[t]{.05\textwidth}{\vspace{-.3\textwidth}(a)}
\parbox[t]{.4\textwidth}{
\includegraphics[width=.3\textwidth]{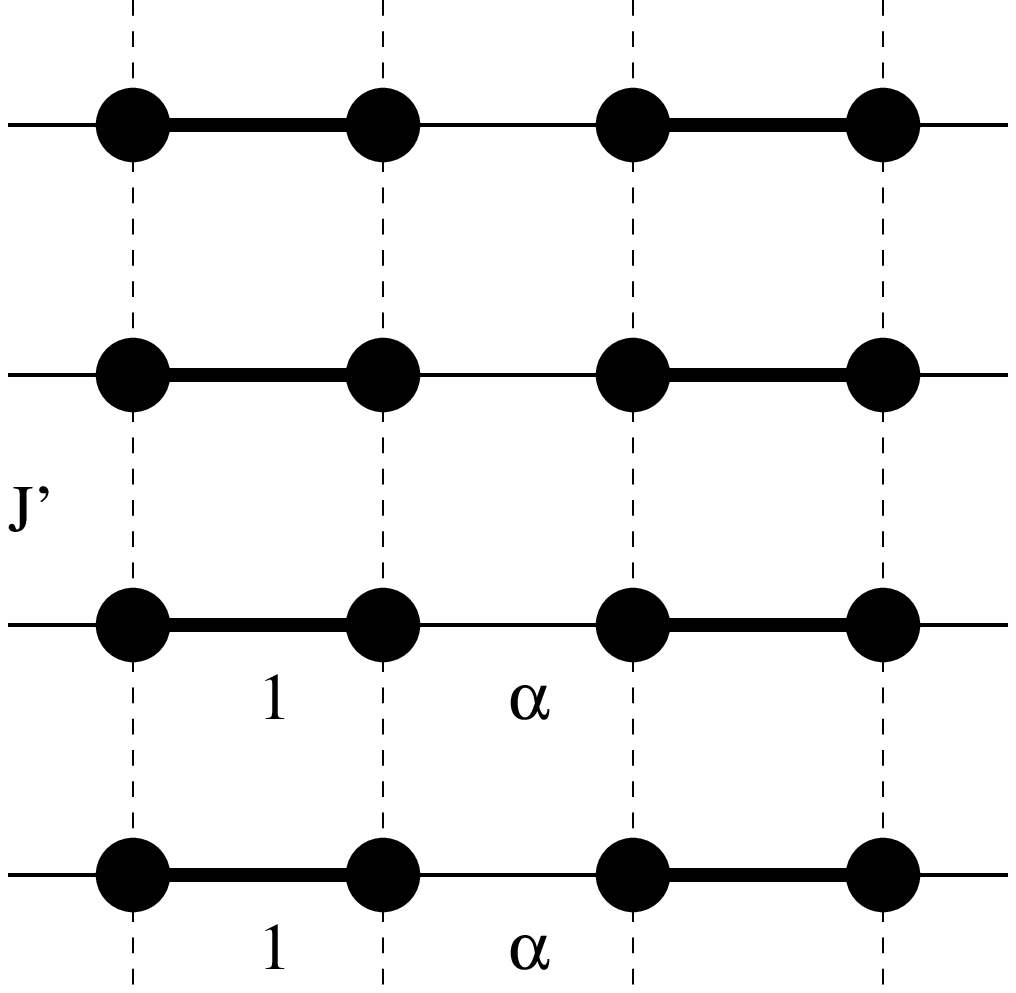}}
\parbox[t]{.05\textwidth}{\vspace{-.3\textwidth}(b)}
\parbox[t]{.4\textwidth}{
\includegraphics[width=.3\textwidth]{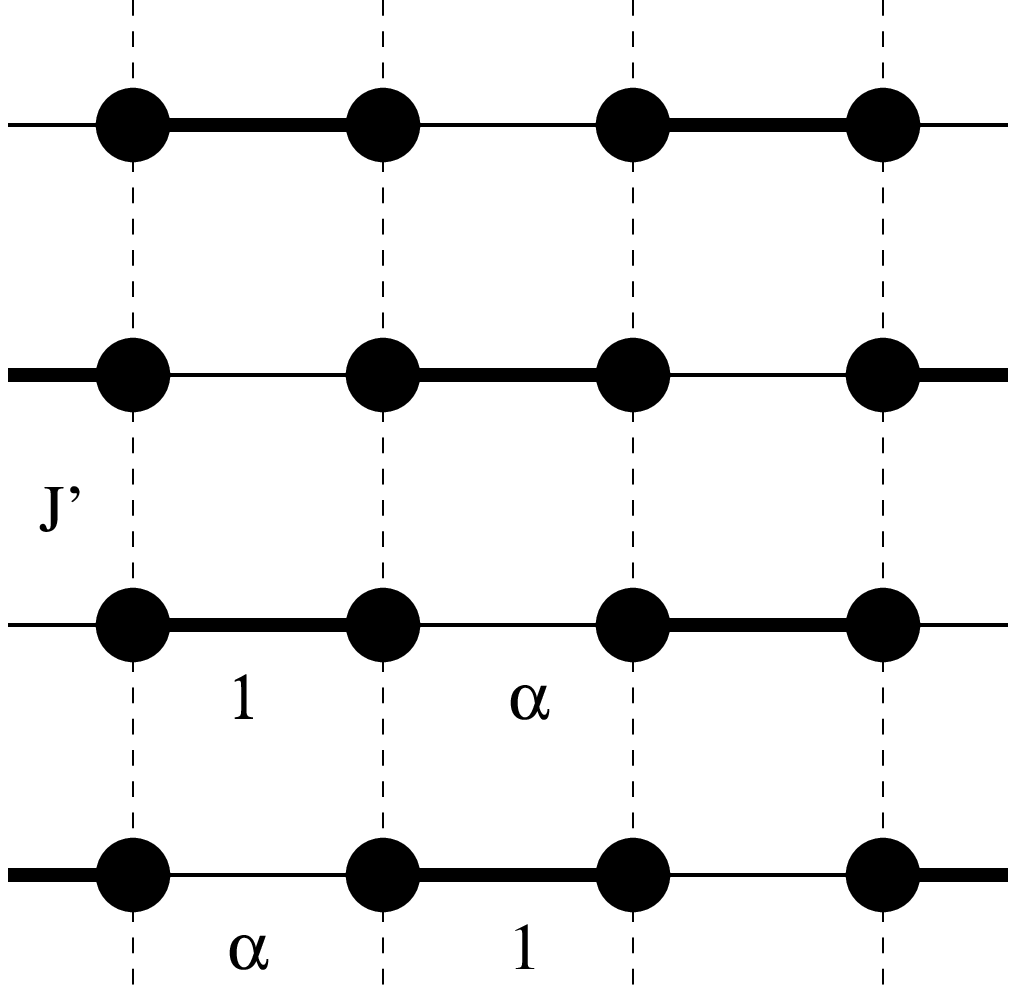}}
\caption{Coupled dimerized chains on a square lattice with
(a) columnar dimerization and (b) staggered dimerization}
\label{model}
\end{figure}

The ground state of a 1D bond-alternated HAF's
with spin magnitude $S=1$
has been investigated extensively~\cite{affleck}. There are
two gapped ground states, the Haldane phase
and the dimer phase, between which a quantum phase
transition of the Gaussian universality class occurs.
%Recently the critical point was determined precisely~\cite{nakamura}.
We study the ground state of coupled $S=1$ bond-alternated spin chains
on a square lattice. There are two ways of coupling such chains,
whether we place the stronger bonds on parallel positions
between the neighboring chains or place them in a zig-zag way.
The arrangements of bonds
are shown in Fig.~\ref{model}.
Hereafter we will refer to the former lattice as the square lattice
with columnar dimerization and the latter as
that with the
staggered dimerization.
The Hamiltonian for the model with columnar dimerization
is written as follows.
\begin{equation}
{\cal H}=J\sum_{i,j}\left(
{\bf S}_{2i-1,j}\cdot{\bf S}_{2i,j}
+\alpha{\bf S}_{2i,j}\cdot{\bf S}_{2i+1,j}\right)
+J'\sum_{i,j}{\bf S}_{i,j}\cdot{\bf S}_{i,j+1}
\end{equation}
The spin operator ${\bf S}_{i,j}$ has magnitude $|{\bf S}|=1$
and $i,j$ denote the points on a square lattice. We consider
only the nearest neighbor coupling.
We set the stronger intrachain coupling
unity and the antiferromagnetic interchain
coupling $J'$ positive.
The strength of bond alternation is
parametrized by $\alpha$ ($0\le\alpha\le 1$).
The Hamiltonian of the model with staggered dimerization
is similarly defined. We set the $x$ axis parallel
to the chains.

There are some previous works
on coupled uniform chains~\cite{sakai,tasaki,koga}
and weakly coupled bond-alternated chains on a square 
lattice with columnar dimerization~\cite{koga}.
It is known that the models have gapped ground states
if the interchain coupling is weak enough
and quantum phase transitions occur
by tuning the interchain coupling.
In our study, by using the continuous-time
quantum Monte Carlo loop algorithm
with the subspin-symmetrization
technique~\cite{todo}, we determine the phase
boundaries accurately and discuss the universality class
of the quantum phase transition. As our method is
numerically exact and non-perturbative, we can derive
the ground-state phase diagram parametrized by $\alpha$ and $J'$
over the whole region.

\section{Simulations and Finite-Size Scaling}
\label{method}

\subsection{Details of Simulations}
Let us denote the size of the lattice simulated
by $L_x$, $L_y$
and the temperature $T=1/\beta$.
By the Suzuki-Trotter decomposition,
we map the original quantum system
on $L_x\times L_y$ lattice to a classical system
on $L_x\times L_y \times \beta$ spacetime by adding
the imaginary-time axis (denoted by $\tau$)
with the length $\beta$.
The aspect ratios
$L_x/L_y$, $L_x/\beta$,
are fixed. Fixing the ratios between the spatial
size and the imaginary-time size is based
on the assumption of Lorentz invariance~\cite{chakravarty},
namely, the dynamical critical exponent be equal to unity.
The sizes of the simulated systems and inverse temperatures
are up to $L_x\times L_y\sim 10^3$ sites and $\beta\sim 10^{2}$,
with $10^3$ Monte Carlo steps used for thermalization and
$10^4$ for measurement. The latter are cut into
20 bins from which we obtain averages and variances
as estimates of observables and their statistical errors.

We calculate the staggered susceptibility $\chi(\pi)$,
correlation lengths $\xi_x$, $\xi_y$ and the excitation gap $\Delta$.
Correlation lengths are calculated from the second moment
of correlation functions
and the gap is obtained as a reciprocal number of $\xi_{\tau}$.
By the finite-size scaling (FSS) of these observables,
we determine critical points and
exponents in the thermodynamic limit $L_x,L_y\rightarrow\infty$
and the ground state limit $\beta\rightarrow\infty$.
We obtain critical exponents of the staggered susceptibility $\gamma$
and the correlation length $\nu$,
by which the critical behavior of
$\chi(\pi)$ and $\xi_{d}$ are described as $\chi(\pi)\sim t^{-\gamma}$ and
$\xi_d \sim  t^{-\nu}$, where $t$ is the distance from the critical point.
These are sufficient
to give other exponents with the scaling relations.

\subsection{Determination of critical points and exponents}

\begin{figure}
\centering
\parbox[t]{.05\textwidth}{\vspace{-.3\textwidth}(a)}
\parbox[t]{.42\textwidth}{\includegraphics[width=.4\textwidth]{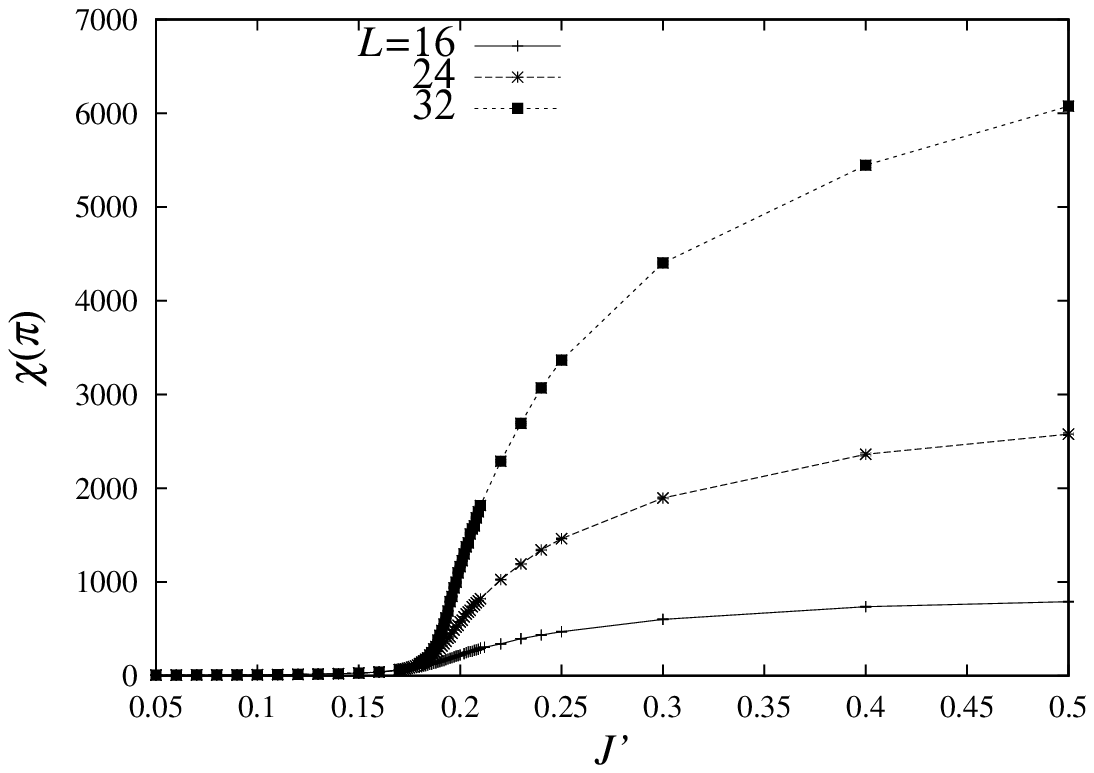}}
\parbox[t]{.05\textwidth}{\vspace{-.3\textwidth}(b)}
\parbox[t]{.42\textwidth}{\includegraphics[width=.4\textwidth]{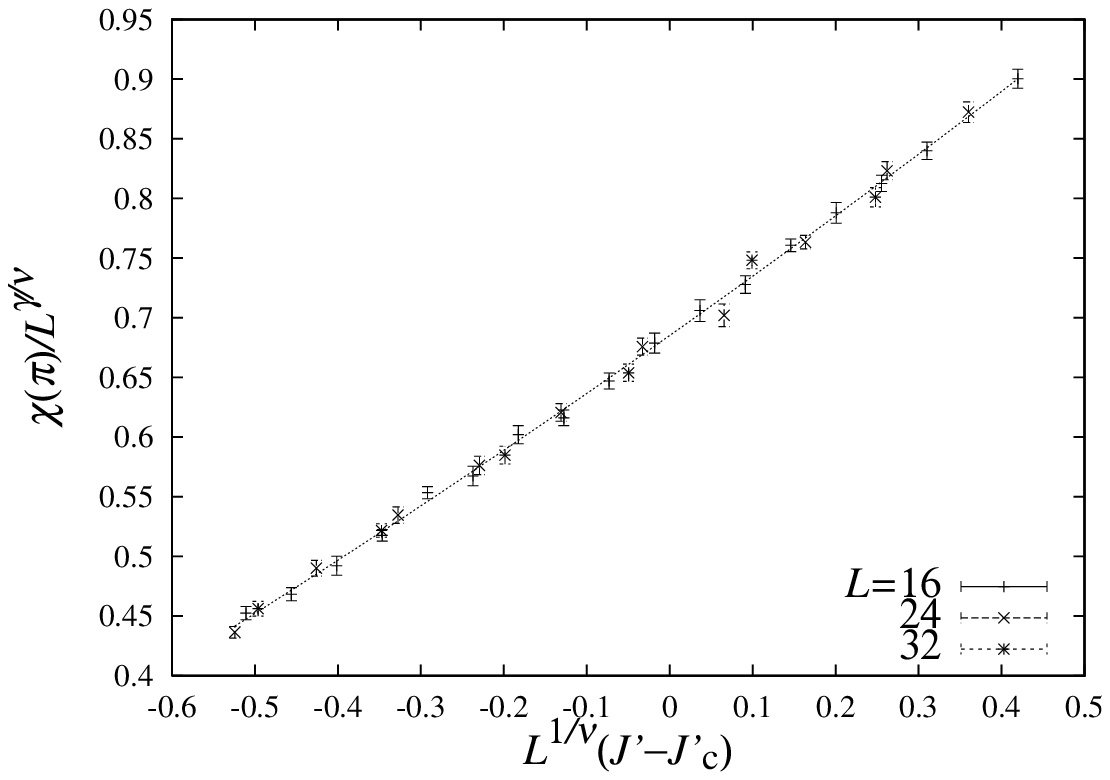}}
\caption{(a) Raw data plot of the staggered susceptibility
and (b) its finite-size scaling plot, with $L$ denoting the
system size simulated}
\label{fss_2D}
\end{figure}

We fit the behavior of observables near critical points
into the FSS formulae which are written as
$\xi_d  =  L f_d(t L^{1/\nu})$ and
$\chi(\pi)  =  L^{\gamma/\nu} g(t L^{1/\nu})$,
where $L$ is the linear system size and $f_d$, $g$ are
polynomials. We take terms of the polynomial
up to the second order. Here we describe how the critical
point and exponents are determined in the ground state
of staggeredly coupled bond alternating chains
with $\alpha=0.1$. The raw data of the staggered susceptibility
$\chi_{\rm s}$ are plotted in Fig.~\ref{fss_2D} (a)
and its FSS is shown in Fig.~\ref{fss_2D} (b).
Data with $L_x=L_y=\beta=16$, $24$, and $32$ are used.
As seen in Fig.~\ref{fss_2D}, the data near the critical point are scaled
quite well by choosing $J'_{\rm c}$, $\nu$ and $\gamma$ as
0.1943(4), 0.69(2) and 1.4(1), respectively.
These exponents coincide with those of 3D classical Heisenberg models
$\nu =  0.7048(30)$ and $\gamma  =  1.3873(85)$~\cite{chen}
within numerical accuracy. It is thus confirmed that
the universality class of the quantum phase transition
of the 2D quantum Heisenberg model
belongs to that of 3D classical Heisenberg models.
The FSS on correlation lengths gives consistent results.

Other critical points are determined in the same way
to yield the phase boundary over the whole parameter region.

\begin{figure}
\centering
\parbox[t]{.05\textwidth}{\vspace{-.3\textwidth}(a)}
\parbox[t]{.42\textwidth}{\includegraphics[width=.4\textwidth]{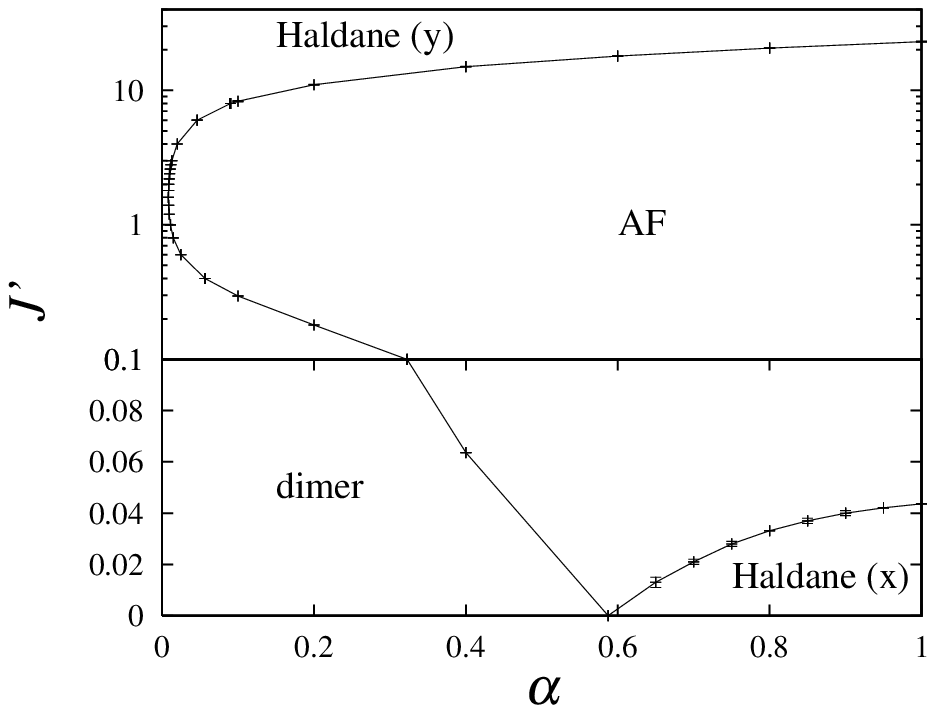}}
\parbox[t]{.05\textwidth}{\vspace{-.3\textwidth}(b)}
\parbox[t]{.42\textwidth}{\includegraphics[width=.4\textwidth]{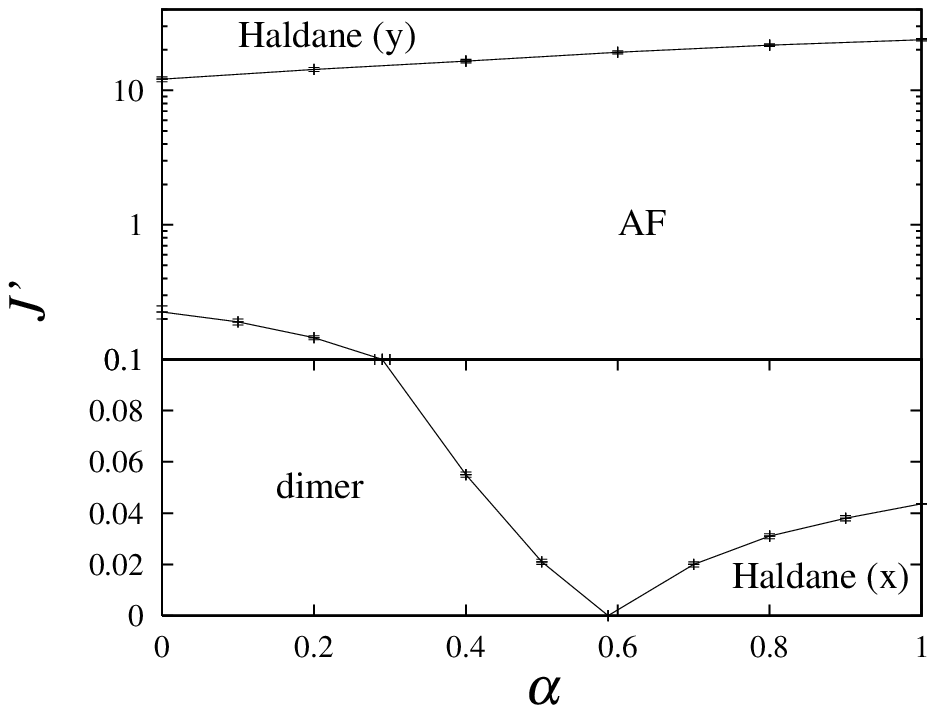}}
\caption{Ground-state phase diagrams of the model on a square
lattice with (a) columnar dimerization and (b) staggered dimerization}
\label{phase_diagram}
\end{figure}

\section{Results and Discussions}
\label{results}

In Fig.~\ref{phase_diagram}, we present the ground-state phase
diagrams for columnar and staggered dimerization.  First of all, both
of them are quite similar with each other for large $\alpha$.
Actually these two models are identical on the line with $\alpha = 1$.
On this line, there are three points which have been investigated in
detail so far.  The 2D isotropic HAF at $(\alpha,J')=(1,1)$ has a
gapless gound state with finite staggered magnetization.  The AF
phase, which includes the isotropic point, occupies a large area in
the phase diagram.  On the other hand, the system consists of
decoupled Haldane chains parallel to the $x$ ($y$) axis in the $J'=0$
($J' \rightarrow \infty$) limit.  The finite excitation gap (Haldane
gap) observed at $J'=0$ and $\infty$ survives even at finite $J'$.  We
refer to these two gapped phases as Haldane (x) and Haldane (y)
phases, respectively.

Now we take a more detailed look on the models with columnar
dimerization.  At $(\alpha,J')=(0,0)$, the system consists of
decoupled dimers and therefore has a spin-gapped gound state (dimer
phase).  This phase also extends to finite $\alpha$ and $J'$.  Most
striking feature of the phase diagram shown in Fig.~1.3(a) is that the
dimer phase around $(\alpha,J')=(0,0)$ and the Haldane (y) phase at
large $J'$ are actually the identical phase.  Since the boundary of
the AF phase does not touch the $\alpha = 0$ line, there is no
critical point between the dimer and Haldane (y) phases.  Especially
it should be pointed out that the $\alpha = 0$ line corresponds to the
two-leg ladders, which always has a spin-gapped gound state
irrespectively of the strength of rung coupling~\cite{todo2}.

Furthermore, the Haldane (x) and Haldane (y) phases are also shown to
be identical by considering the bond alternation in the $y$
direction~\cite{mm}.  Thus in the ground states of Q1D models, all of
the gapped phases are identical.  On the other hand, in a strictly 1D
chain, the dimer phase and the Haldane phase are definitively
distinguished in terms of the topological hidden order measured by the
string-order parameter~\cite{denNijs}, which is non-zero only in the
Haldane phase.  It should be emphasized that once we have a finite
interchain coupling the string-order parameter vanishes even in the
Haldane (x) phase.  In this sense the line $J'=0$, which represents
strictly 1D chains, is singular in the phase diagram.

The ground-state phase diagram of the model with staggered
dimerization is topologically different from the columnar one.
Particularly the AF phase extends onto the $\alpha=0$ line, as the
lattice remains connected two-dimensionally even at $\alpha=0$ (as
long as $J'$ is finite).  Thus in the phase diagram three spin-gapped
phases (Haldane (x), Haldane (y) and dimer) are separated by the AF
phase.  It is of great interest to pursue phase diagrams in the
presence of the bond alternation in the $y$ direction
both in columnar and staggered ways and see the
topology of these gapped phases in the extended parameter space.

% BibTeX users please use
% \bibliographystyle{}
% \bibliography{}
%
% Non-BibTeX users please follow the syntax
% the syntax of "referenc.tex" for your own citations

%%%%%%%%%%%%%%%%%%%%%%%%%%%%%%%%%%%%%%%%%%%%%%%%%%%%%%%%%%%%%%%%%%%%%%  }

%%%%%%%%%%%%%%%%%%%%%%%%%%%%%%%%%%%%%%%%%%%%%%%%%%%%%%%%%%%%%%%%%%%%%%

\printindex
\end{document}